\documentclass{appolb}
\usepackage{graphicx,amsmath,amssymb}
\usepackage[colorlinks,linkcolor=blue,citecolor=red]{hyperref}

\begin{document}

\title{Far-from-equilibrium attractor in non-conformal plasmas%
\thanks{Presented at Quark Matter 2022}%
}
\author{Sunil Jaiswal, Subrata Pal\\[-4mm]
\address{Department of Nuclear and Atomic Physics, Tata Institute of Fundamental Research, Mumbai 400005, India}\\[2mm]
Chandrodoy Chattopadhyay\\[-4mm]
\address{Department of Physics, North Carolina State University, Raleigh, NC 27695, USA}\\[2mm]
Lipei Du\\[-4mm]
\address{Department of Physics, McGill University, Montreal, QC H3A 2T8, Canada}\\[2mm]
Ulrich Heinz\\[-4mm]
\address{Department of Physics, The Ohio State University, Columbus, OH 43210, USA}\\[-2mm]
}

\maketitle
\begin{abstract}
We study the far-off-equilibrium dynamics of a Bjorken expanding non-conformal system within kinetic theory and hydrodynamics. We show that, in contrast to the conformal case, neither shear nor bulk viscous pressure relax quickly to a non-equilibrium attractor. In kinetic theory an early-time, far-from-equilibrium attractor exists for the scaled longitudinal pressure, driven by the rapid longitudinal expansion of the medium. Second-order dissipative hydrodynamics fails to accurately describe this attractor, but a modified anisotropic hydrodynamic formulation reproduces it and provides excellent agreement with kinetic theory.
\end{abstract}


\section{Introduction}

In recent years, several studies comparing higher-order hydrodynamic theories to conformal kinetic theory in boost-invariant flow profiles have revealed a surprising success of hydrodynamics in providing a near-accurate description of the system's macroscopic dynamics even when the medium is very far from local equilibrium. An important feature that emerged from these studies is the existence of a far-from-equilibrium attractor in hydrodynamic theories \cite{Heller:2015dha} to which various initializations of viscous stresses decay via power law at large Knudsen-numbers \cite{Jaiswal:2019cju, Kurkela:2019set}. However, almost all these comparisons of hydrodynamics with kinetic theory have focused on conformal systems with highly symmetric expansion profiles. Here we investigate the domain of applicability of second-order {\it non-conformal} hydrodynamics, by comparing it with kinetic theory for systems undergoing (0+1) dimensional expansion with Bjorken symmetry \cite{Chattopadhyay:2021ive, Jaiswal:2021uvv}.

\section{Kinetic theory}

We consider the evolution of the single particle distribution function $f(x,p)$ satisfying Bjorken symmetries, described by the Boltzmann equation with a collision term in relaxation time approximation (RTA):
\begin{equation}
\label{RTA}
	\frac{\partial f}{\partial \tau} = - \frac{f - f_{\rm eq}}{\tau_R(\tau)}.
\end{equation}
$f_{\rm eq}$ represents the equilibrium distribution function. We parametrize the relaxation time as $\tau_R = 5\,C/T$, where $T$ is the temperature and $C$ is a unitless constant. The above kinetic equation can be solved exactly \cite{Florkowski:2014sfa}, and appropriate moments of the distribution function give the exact evolution of the hydrodynamic quantities. The initial distribution function is parametrized to allow for large initial bulk and shear stresses \cite{Chattopadhyay:2021ive, Jaiswal:2021uvv}.

\subsection{Kinetic bounds on viscous stresses}

Bjorken symmetry dictates the energy momentum tensor to be diagonal, $T^{\mu\nu} = {\rm diag} (\epsilon, P_T, P_T, P_L)$, where $\epsilon$, $P_T$ and $P_L$ are the energy density and effective transverse and longitudinal pressures, respectively. The latter can be expressed in terms of the equilibrium pressure ($P$), bulk viscous pressure ($\Pi$), and a single independent shear stress tensor component $\pi \equiv -\tau^2 \pi^{\eta \eta}$: $P_T = P{+}\Pi{+}\pi/2$ and $P_L = P{+}\Pi{-}\pi$. Positivity of distribution function imposes in kinetic theory the following bounds on the normalized viscous stresses $\bar\pi\equiv \pi/P$ and $\bar\Pi\equiv \Pi/P$ \cite{Chattopadhyay:2021ive}:
\begin{equation}
\label{B2}
	\bar\Pi + \frac{1}{2}\bar\pi \geq -1, \quad  
	\bar\Pi - \bar\pi \geq -1, \quad 
	\bar\Pi \geq - 1, \quad 
	\bar\Pi \leq \frac{\epsilon}{3P} - 1. 
\end{equation}
Solutions of the kinetic equation satisfy these bounds at all times.

\begin{figure}[t]
\centerline{%
\includegraphics[width=0.95\textwidth]{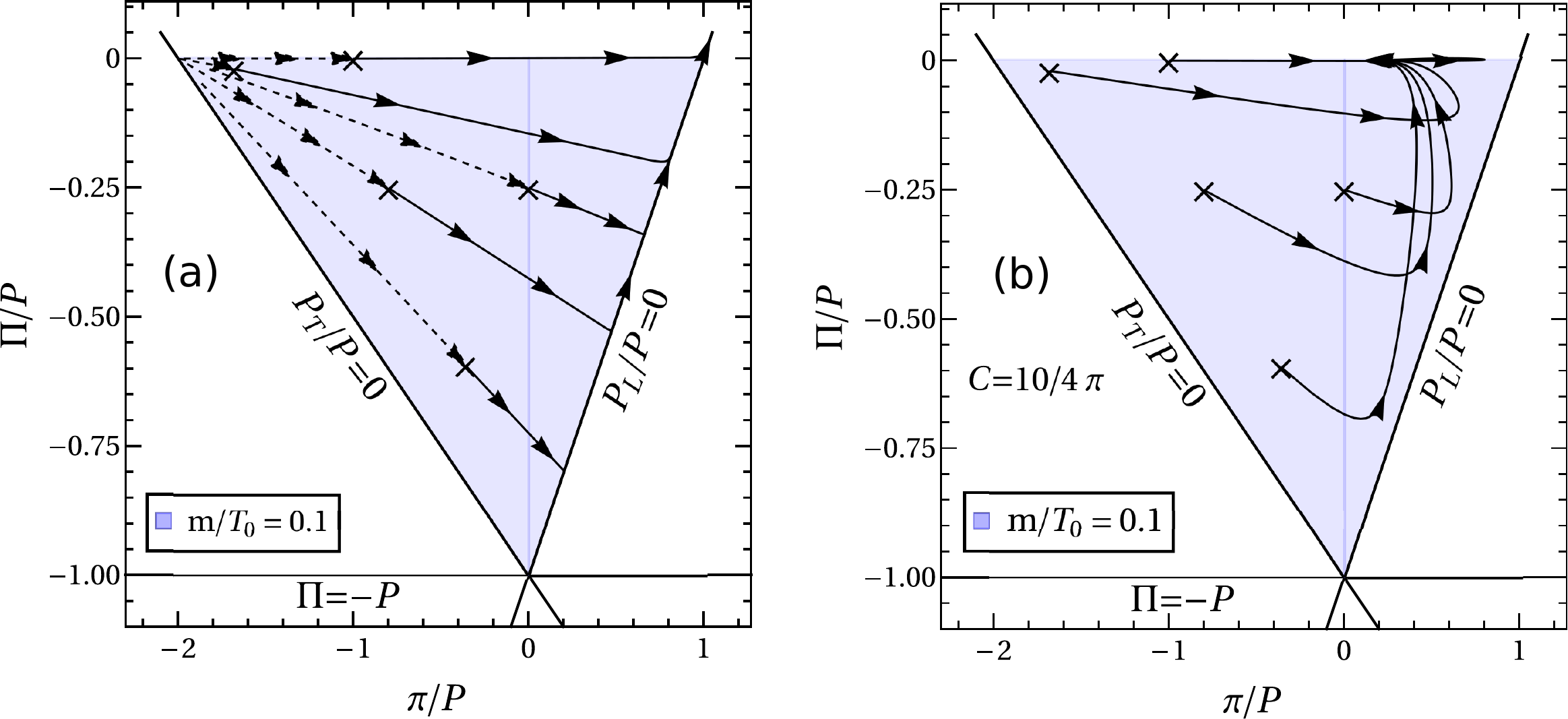}}
\caption{Evolution of viscous stresses in the free-streaming limit (a) and including microscopic interactions (b). Figures adapted from Refs.~\cite{Chattopadhyay:2021ive, Jaiswal:2021uvv}.}
\label{Fig1}
\vspace*{-4mm}
\end{figure}

\vspace{-.1cm}
\subsection{Free-streaming evolution}

In Bjorken flow the early time dynamics of the medium is governed by fast longitudinal expansion and approximately free-streaming. The free-streaming solution of Eq.~\eqref{RTA} is simply $f_\mathrm{fs}(\tau; p_T, p^z) = f_\mathrm{in}(p_T, p^z (\tau/\tau_0))$, i.e., the distribution function becomes sharply peaked around $p^z=0$ as time increases. Some free-streaming trajectories in the $\bar\pi{-}\bar\Pi$ plane are shown in Fig.~\ref{Fig1}a; the shaded blue region represents the bounds mentioned in Eq.~\eqref{B2}.  The black crosses indicate different initializations with $m = 50$\,MeV and $T_0 = 500$\,MeV at initial time $\tau_0 = 0.1$\,fm/$c$, and arrows indicate the direction of time. All trajectories are seen to move towards the line $P_L=0$ which acts like an attractive fixed line. Under backward evolution, represented by the dashed lines, all trajectories are seen to merge at the point $(\Pi/P=0, \pi/P=-2)$ as $\tau\to 0$; this point is thus a repulsive dynamical fixed point.

\section{Dynamics at finite Knudsen number}

Microscopic interactions force the medium to depart from free-streaming dynamics and drive it towards local momentum isotropy. Their effect on the expansion trajectories is shown
in Fig.~\ref{Fig1}b. Microscopic collisions build up longitudinal pressure and thus push the trajectories away from the $P_L=0$ line. Eventually the system thermalizes locally and, after reaching the Navier-Stokes limit, converges to the thermal equilibrium point $\bar\pi=\bar\Pi=0$.

\begin{figure}[t]
\centerline{%
\includegraphics[width=\textwidth]{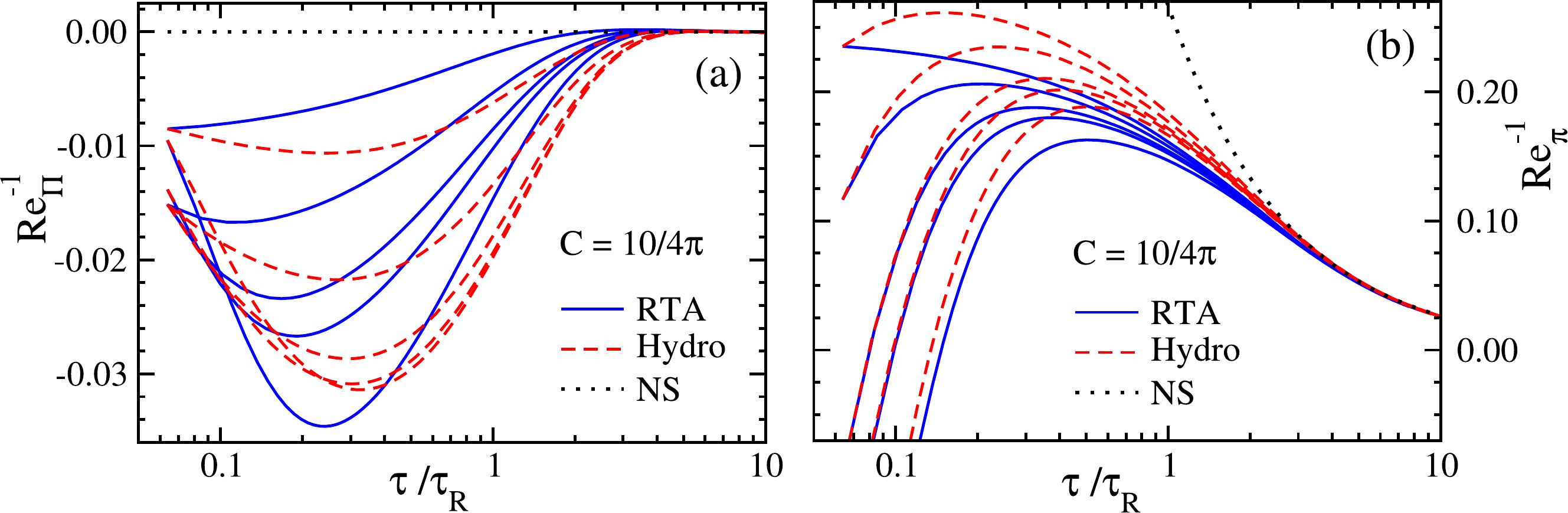}}
\caption{Evolution of the (a) bulk and (b) shear inverse Reynolds numbers. Figures adapted from Refs.~\cite{Chattopadhyay:2021ive, Jaiswal:2021uvv}.}
\label{Fig2}
\vspace*{-4mm}
\end{figure}

Fig.~\ref{Fig2} shows the evolution of the bulk and shear inverse Reynolds numbers, Re$^{-1}_{\Pi} \equiv \Pi/(\epsilon{+}P)$ and  Re$^{-1}_{\pi} \equiv \pi/(\epsilon{+}P)$, as functions of the scaled time $\bar{\tau} \equiv \tau/\tau_R$, for a microscopic interaction strength given by $C{\,=\,}10/4\pi$. Blue solid curves are solutions of the RTA Boltzmann equation and red dashed ones represents solutions of second-order Chapman-Enskog hydrodynamics \cite{Jaiswal:2014isa}. The black dotted curve is the first-order Navier-Stokes (NS) solution.

One naively expects to recover the attractor structure of conformal systems for evolution with small values of bulk viscous pressures. However, the shear stress trajectories in Fig.~\ref{Fig2}b as seen to converge only at $\bar\tau{\,\gtrsim\,}2$ (magnitudes of the initial bulk viscous pressure are small as can be seen in Fig.~\ref{Fig2}a). This  contrasts sharply with the pattern observed in conformal systems where different initializations of shear stresses rapidly approach an early-time attractor on a much shorter time scale that is controlled by the initialization time $\tau_0$ \cite{Jaiswal:2019cju, Kurkela:2019set}. For large initial bulk stress the convergence of trajectories is delayed even further \cite{Chattopadhyay:2021ive, Jaiswal:2021uvv}. Therefore, we conclude that in non-conformal fluids there is no evidence of early-time attractors for the shear and bulk viscous stresses.

The absence of far-off-equilibrium attractors for the viscous stresses begs the question whether such an attractor exists at all in non-conformal systems. We find that an early-time attractor does manifest in the evolution of the effective longitudinal pressure. The key to finding it is the realization that for Bjorken flow, due to the divergence of the expansion rate at $\tau\to0$, the early time dynamics is approximately free-streaming, and the line $P_L{\,=\,}0$ acts as an attractor for this free-streaming medium (see Fig.~\ref{Fig1}a).

\begin{figure}[t]
\centerline{%
\includegraphics[width=\textwidth]{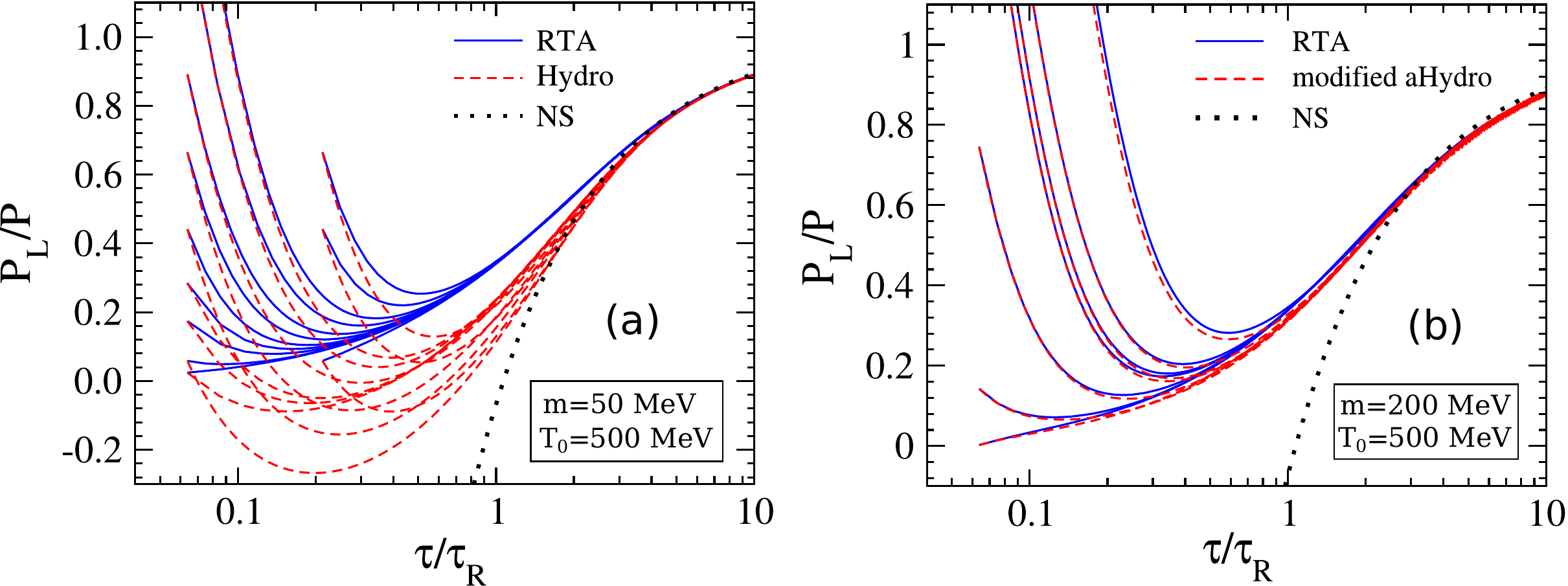}}
\caption{Evolution of the scaled longitudinal pressure. Figures adapted from~\cite{Chattopadhyay:2021ive, Jaiswal:2021uvv}.}
\label{Fig3}
\vspace*{-3mm}
\end{figure}

Fig.~\ref{Fig3} shows the evolution of the scaled longitudinal pressure $P_L/P$ as a function of the scaled time $\bar\tau$. The initial conditions considered in Fig.~\ref{Fig3}a correspond to those of Fig.~\ref{Fig2}, amended by additional initial conditions with vanishing shear stress. The interaction strength parameter has been changed to $C=3/4\pi$ (shorter relaxation times). Kinetic theory solutions (blue solid lines) are seen to join already at times $\bar\tau{\,<\,}1$ a universal attractor that starts from $P_L/P{\,\approx\,}0$ at $\bar\tau_0{\,\to\,}0$. As the system isotropizes, this universal curve approaches unity, joining the first-order Navier-Stokes solution (black dotted curve) at $\bar\tau{\,\gtrsim\,}4$. The same features are seen in Fig.~\ref{Fig3}b where we have considered different initial conditions with $m=200$\,MeV. We conclude that only $P_L{\,=\,}P{+}\Pi{-}\pi$ has a far-from-equilibrium attractor, driven by the approximate free-streaming dynamics of Bjorken flow at early times. Only in conformal systems, where $P_L$ and $\pi$ describe the same physics ($P_L/P=1-\pi/P$), they also share this attractor.

The non-conformal hydrodynamic trajectories in Fig.~\ref{Fig3}a (red dashed lines) do not exhibit a universal early-time attractor; universality is seen only at $\bar\tau{\,\gtrsim\,}4$ when they merge with the NS attractor. Clearly, second-order non-conformal hydrodynamics is not an accurate approximation of the underlying kinetic theory before $\bar\tau{\,\simeq\,}3$.

\vspace*{-2mm}
\section{Anisotropic hydrodynamics}\label{aHydro}

We now come to the dashed red lines in Fig.~\ref{Fig3}b. In the standard procedure, hydrodynamic equations are derived from kinetic theory by expanding the phase-space distribution function $f(x,p)$ around a locally isotropic equilibrium distribution. However, for Bjorken flow this distribution, and its leading-order corrections, do not follow the rapid shrinking of the $p_z$ distribution caused by the rapid expansion rate at early times. As a result, hydrodynamics fails to reproduce the early-time behavior. 

To address this shortcoming {\it anisotropic hydrodynamics} (aHydro) was introduced~\cite{Florkowski:2010cf, Martinez:2010sc, Bazow:2013ifa}. However, the standard derivation of aHydro, which is based on an expansion of the distribution function around an ellipsoidally deformed distribution of Romatschke-Strickland form \cite{Romatschke:2003ms}, does not allow to simultaneously generate large temperatures ($T \gg m$) and large bulk viscous pressures $\Pi/P\simeq-1$ \cite{Jaiswal:2021uvv}. This problem can be circumvented by considering a modified ansatz for the leading-order anisotropic distribution \cite{Jaiswal:2021uvv}, with an additional non-equilibrium fugacity factor:
\begin{equation} 
\label{modifed_aHydro_ansatz}
    f \approx \tilde{f}_a = \frac{1}{\alpha(\tau)} \exp\Biggl(-\frac{\sqrt{p_T^2 + (1{+}\xi(\tau)) w^2/\tau^2 + m^2}}{\Lambda(\tau)} \Biggr). 
\end{equation}

In Fig.~\ref{Fig3}b we compare the evolution described by the resulting modified aHydro equations with the exact kinetic result, for the scaled longitudinal pressure (comparisons for other hydrodynamic quantities can be found in \cite{Jaiswal:2021uvv}). The modified aHydro solutions (red dashed lines) are seen to be in excellent agreement with the exact kinetic results (blue solid lines). Modified aHydro also reproduces the rapid convergence of solutions with arbitrary initial conditions onto an early-time far-from-equilibrium attractor (the lowermost blue solid curve) at $\bar\tau{\,<\,}1$, long before they merge with the late-time Navier-Stokes attractor (black dotted curve) at $\bar\tau{\,\gtrsim\,}3$.

\section{Conclusions}

We studied the expansion of a non-conformal system whose microscopic dynamics is governed by the RTA Boltzmann equation. In kinetic theory we demonstrated that, different from conformal systems, only the longitudinal pressure features an early-time, far-off-equilibrium attractor. It is governed by the rapid, approximately free-streaming expansion of the medium at early times, and its existence relies on the fact that for Bjorken flow at early times the expansion rate hugely exceeds the scattering rate. Shear and bulk viscous stresses do not feature a far-off-equilibrium attractor. Their early-time dynamics is characterized by strong bulk-shear coupling effects which depend sensitively on initial conditions. Only in conformal systems, where the shear stress and longitudinal pressure describe identical physics, do $P_L$ and $\pi$ share an early-time attractor.

We also found that for non-conformal systems standard dissipative hydrodynamics is unable to describe the early-time dynamics, in particular the far-from-equilibrium attractor for $P_L$. However, a modified version of anisotropic hydrodynamics reproduces it well.

\bibliographystyle{h-physrev}
\bibliography{reference}

\end{document}